\documentclass[12pt,a4paper,final]{article}
\usepackage[utf8]{inputenc}
\usepackage{amsmath}
\usepackage[english]{babel}
\usepackage{amsfonts}
\usepackage{setspace}
\usepackage{hyperref}
\usepackage{amssymb}
\usepackage{dsfont}
\usepackage{makeidx}
\addto{\captionsenglish}{}
\date{}
\usepackage{graphicx}
\usepackage[left=2.5cm, right=2.5cm]{geometry}
\author{$FAICAL\quad BARZI^{1,2}$\\\small faical.barzi@edu.uiz.ac.ma\\\\\small $^1$Centre Regional des Métiers de l'Education et de la Formation\\\small CRMEF-Marrakech-Safi, Morocco\\\small $^2$Ibn Zohr University-Agadir, Morocco}
\title{\Huge On Complex numbers in quantum Mechanics }

\begin{document}
\maketitle

\begin{abstract}
	We look at the fundamental use of complex numbers in Quantum Mechanics (QM). A review of some of the most popular reasons given in the literature to support the necessity of the complex formalism,  We add some insight by invoking others. This short study is aimed at strengthening the delicate pedagogical endeavor of explaining Why Quantum mechanics needs complex numbers.

\end{abstract}

\section*{INTRODUCTION}
The effectiveness of mathematics in describing physical systems is a long standing puzzle of natural sciences\cite{mat1}\cite{mat2}. Suggestive work proposes a deep intertwining between properties of physical and mathematical systems\cite{intertwin}. In this regard, the important question of why complex numbers appear in Quantum Mechanics ? arises. Complex numbers were discovered by $16^{th}$ century mathematicians in the course of their attempt to solve cubic equations\cite{hist}. They went to find many application is different branches of pure mathematics, classical physics and engineering as basis for powerful mathematical methods. However, the advent of QM raised the status of complex numbers to that of real numbers, sometimes with direct physical manifestations\cite{aharo}\cite{res th}. The founders of QM were compelled to use complex numbers but remained puzzled by this seemingly unnecessary appearance\cite{schrodi}. Many attempts to reformulate QM with real numbers are proposed\cite{realQM}\cite{realQM1}\cite{realQM2}, but the equivalence of such reformulations to complex QM is still to be proven\cite{contro}. On a pedagogical level, assimilating the role and necessity of complex quantities in QM ease the acceptance of newcomers to QM of this intrusion of imaginary numbers and at the same time deepens the understanding of the mathematical formalism of the theory. Many authors\cite{auth}\cite{auth1}\cite{auth2} presented their take on the necessity and utility of complex numbers in QM, our task here is to review their presentation and provide new insight to this fundamental topic and to dispel a common misconception that it's the Schrodinger's equation that introduces complex numbers into QM\cite{misc} rather than being a mere consequence of their necessity as we show below.
\section{Correspondence principal}
Dirac\cite{Dirac} introduced the correspondence between Poisson brackets defined as : 
\begin{equation}\label{key}
	 {\left\lbrace A,B\right\rbrace=\sum\left[ \frac{\partial A}{\partial q_i}\frac{\partial B}{\partial p_i}-\frac{\partial A}{\partial p_i}\frac{\partial B}{\partial q_i}\right]} 
\end{equation} and the commutator $\left[A,B\right]=AB-BA$, of two observables $A$ and $B$. The mathematical resemblance between the two objects led Dirac to formulate his form of the correspondence principal which states that statements in QM which contain a commutator \textit{correspond} to statements in classical mechanics where the commutator is replaced by a Poisson bracket multiplied by i$\hbar$. In the classical limit $\hbar\longrightarrow0$, we expect all observables to be compatible and commute with each other. Thus a priori $\left[A,B\right]=\eta\left\lbrace A,B\right\rbrace=\eta\left\lbrace A,B\right\rbrace \mathds{1}$, where $\mathds{1}$ is the identity operator and $\eta$ must verify the following conditions.

\begin{enumerate}
	\item the classical limit, $\eta\longrightarrow0$.
	\item In terms of dimensions, $ \eta $ must dimension of [q][p], that is dimension of action expressed in $J.s$. Thus $\eta=\alpha\hbar$, where alpha is a number.
	\item Poisson brackets are real numbers :  $ \left\lbrace A,B\right\rbrace=\left\lbrace A,B\right\rbrace^*=\left\lbrace A,B\right\rbrace^T$
		\begin{align}
		& \left\lbrace A,B\right\rbrace=	\frac{\left[A,B\right]}{\alpha\hbar}\label{first1}\\
		& 	\left\lbrace A,B\right\rbrace^T=\frac{\left[A,B\right]^T}{\alpha\hbar}\\
		& 	\left\lbrace A,B\right\rbrace=\frac{\left[B^T,A^T\right]}{\alpha\hbar}\\ 
		& 	\left\lbrace A,B\right\rbrace=-\frac{\left[A^T,B^T\right]}{\alpha\hbar}\label{last1}	
	\end{align}
	By comparing (\ref{first1}) and (\ref{last1}), we see that :
	\begin{equation}\label{first2}
		\left[A,B\right]=-\left[A^T,B^T\right]
	\end{equation}
	On the other hand,
	\begin{align}
		& \left\lbrace A,B\right\rbrace=	\frac{\left[A,B\right]}{\alpha\hbar}\label{first}\\
		& 	\left\lbrace A,B\right\rbrace^*=\frac{\left[A^*,B^*\right]}{\alpha^*\hbar}\\
		& 	\left\lbrace A,B\right\rbrace=\frac{\left[A^*,B^*\right]}{\alpha^*\hbar}\label{last}	
	\end{align}
Comparing (\ref{first}) and (\ref{last}), we have :
\begin{equation}\label{last2}
	\alpha^*\left[A,B\right]=\alpha\left[A^*,B^*\right]
\end{equation}

From (\ref{first2}) and (\ref{last2}), we deduce that :

\begin{equation}\label{key}
		\alpha^*\left[A^T,B^T\right]=-\alpha\left[A^*,B^*\right]
\end{equation}
For all observables $A$ and $B$. This condition requires that:
\[
\begin{cases}
\left[A^T,B^T\right]=\left[A^*,B^*\right]\\
\alpha=-\alpha^* 
\end{cases}\]
The first condition implies that A and B are Hermitian operators i.e. $A^*=A^T$, while more importantly, the second says that $\alpha$ must be a pure imaginary number i.e. $\alpha=i\:l$, where $l\in \mathbb{R}$. Thus In QM, the correspondence principal imposes the use of complex numbers and the Hermiticity of observables. 
\end{enumerate}

As an aside, all physical quantities are real and in QM they correspond to eigenvalues of the operator associated with every physical quantity. No complex physical observable is known to date, Let $\xi$ be a physical quantity. Let $\Psi$ be a quantum wave function of a given system then all wave functions form a vector space close under the action of all observables and then $\Phi=\hat{\xi}\Psi$ must also be valid wave function of the system. Since $\hat{\xi}$ is Hermitian, wave functions are complex in general and the vector space is a Hilbert space. Whether the Hermiticity of the Hamiltonian is a necessary or a sufficient condition remains a question to be settled. A complex extension of QM\cite{ext}, suggest that a non-Hermitian observables exhibiting a $\mathcal{CPT}-$symmetry could produce consistent "doubly" complex quantum theories.

\section{Time Evolution Operator}
In QM the time evolution is a continuous process, that can be subdivided indefinitely into smaller elementary steps, each step must be described by a valid time evolution operator. Let $U(t_2,t_1)$ be the time evolution operator of a given system between times $t_1$ and $t_2$, where $t_2>t_1$. For any time $t$, such that $t_1<t<t_2$ we have:
\begin{equation}\label{key}
	U(t_2,t_1)=U(t_2,t)U(t,t_1)
\end{equation}
If we choose $t$ such that $t_2-t=t-t_1$ we can write	  $U(t_2,t)=U(t,t_1)=U(t)$, thus :
\begin{align}\label{key}
	&	U(t_2,t_1)=\{U(t)\}^2\\
	&\implies 	U(t)=\sqrt{U(t_2,t_1)}
\end{align}
The requirement that the square root of any time evolution operator must exist imposes the use of complex numbers whose field is algebraically closed, i.e. Every nth power algebraic equation has precisely $n$ solutions in this field. Of course not just the square roots that must exist but all n-roots of $	U(t_2,t_1)$. This goes to shows also that without complex numbers there could be no time evolution in QM. Schrodinger' equation then must be a complex differential equation if it is to dictate the time evolution in QM.
\section{Spin and Uniformity of Space}
Consider the component of spin of a particle along an arbitrary direction of space noted $\sigma_z$ which has eigenvalue $\pm1$. In the Hilbert space associated with the spin state of the particle we can express $\sigma_z$ as:
\begin{equation}\label{key}
	\sigma_z=\left|z+\right\rangle\left\langle z+ \right|-\left|z-\right\rangle\left\langle z- \right|
\end{equation}
The completeness relation for the spin of the particle reads:
\begin{equation}\label{key}
	\mathds{1}=\left|z+\right\rangle\left\langle z+ \right|+\left|z-\right\rangle\left\langle z- \right|
\end{equation}
The spin components $\sigma_x$ and $\sigma_y$ in the two directions orthogonal to $z$ can be expressed in the base of the projector operators :
\begin{equation}\label{key}
\{\;	\left|z+\right\rangle\left\langle z+ \right|,\left|z-\right\rangle\left\langle z- \right|,\left|z+\right\rangle\left\langle z- \right|,\left|z-\right\rangle\left\langle z+ \right|\;\}
\end{equation}
as:
\begin{align}\label{key}
	&\sigma_x=A\left|z+\right\rangle\left\langle z+ \right|+B\left|z-\right\rangle\left\langle z- \right|+C\left|z+\right\rangle\left\langle z- \right|+D\left|z-\right\rangle\left\langle z+ \right|\\
		&\sigma_y=a\left|z+\right\rangle\left\langle z+ \right|+b\left|z-\right\rangle\left\langle z- \right|+c\left|z+\right\rangle\left\langle z- \right|+d\left|z-\right\rangle\left\langle z+ \right|
\end{align}

Since all directions of space are equivalent and space is uniform\cite{sch1}, $\sigma_x$ and $\sigma_y$ must have the same eigenvalues as $\sigma_z$:
\begin{align}\label{key}
	\sigma_x^2=1\\
\sigma_y^2=1
\end{align}
 thus the requirements of incompatibility of the spin components imposes that:
\begin{align}\label{key}
&	A=a=B=b=0
\end{align}

$\sigma_x$ and $\sigma_y$ reads:
\begin{align}\label{key}
		&\sigma_x=C\left|z+\right\rangle\left\langle z- \right|+D\left|z-\right\rangle\left\langle z+ \right|\\
	&\sigma_y=c\left|z+\right\rangle\left\langle z- \right|+d\left|z-\right\rangle\left\langle z+ \right|
\end{align}
On the other hand, imposing the uniformity of space :
\begin{align}\label{key}
&C\:D=1\\
&c\:d=1\\
&||C||=||D||\\
&||c||=||d||
\end{align} 
The two last conditions comes from the arbitrariness of what we choose to label  $z+$  and $z-$ directions. We see that $c$ and $d$ are solutions of the same equation as $C$ and $D$ Let $C=re^{i\theta_C}$ and  $D=re^{i\theta_D}$ we write:
\begin{equation}\label{key}
	r^2e^{i(\theta_C+\theta_D)}=1
\end{equation}
 Putting $r=1$, we have $\theta_C=-\theta_D[2\pi]$ and $\theta_c=-\theta_d[2\pi]$. Thus, given $\theta$ and $\theta'$, two different angles, such that $\theta\neq\theta'[\pi]$ we have:
\begin{align}\label{key}
	&\sigma_x=\left|z+\right\rangle\left\langle z- \right|+e^{-2i\theta}\left|z-\right\rangle\left\langle z+ \right|\\
	&\sigma_y=\left|z+\right\rangle\left\langle z- \right|+e^{-2i\theta'}\left|z-\right\rangle\left\langle z+ \right|
\end{align}
One of many choices is : $\theta=0$ and $\theta'=\frac{\pi}{2}$. Nothing forces us a priori to make this choice, the more important conclusion is the need of complex numbers to describe the spin of a particle.
\section{Time Reversal Symmetry}
Given a system whose Hamiltonian $H$ is time-independent, its energy is conserved. Such a system must posses a time reversal symmetry\cite{Col} i.e. upon replacing $t\longrightarrow-t$, the system evolve according to the same equation of the motion. Let $U=e^{-iHt}$ be the time evolution of the system, a time reversal changes this operator to $e^{iHt}$ according to :
\begin{equation}\label{t1}
	U_T^\dagger e^{-iHt}U_T=e^{iHt}
\end{equation}
Where $ U_T $ is a priori a (linear) unitary operator representing time reversal symmetry and commutes with the Hamiltonian: 
\begin{align}
	[H,U_T]=0\label{comm}\\
	U_T^\dagger=U_T^{-1}
\end{align} Now, for a infinitesimal time step, $ e^{-iHt}=1-iH\delta t +O(\delta t^2)$, substituting in (\ref{t1}):
\begin{align}
	&	U_T^\dagger H U_T=-H \label{unb}\\
	& 	H U_T=-U_TH\\
	&   	H U_T+U_TH=0
\end{align}

Which imply that $H$ and $U_T$ anti-commute, also from (\ref{unb}) unitary time reversal transforms $H\longrightarrow-H$ which makes the energy spectrum unbounded from below. These contradictions reflect the inconsistency of a unitary operator representation of time reversal symmetry with the laws of QM. The correct way to define a representation of the time reversal symmetry is through an anti-unitary(anti-linear) operator\cite{wig}, say $\Omega_T$,  which verifies $\Omega_T^{^-1}(-i)\Omega_T=i\Omega_T^{^-1}\Omega_T=i$ such that:
\begin{equation}\label{key}
	\Omega_T^{^-1}e^{-iHt}\Omega_T=e^{iHt}
\end{equation}
For an infinitesimal time step:
\begin{align}\label{key}
	&	\Omega_T^{^-1}(-i)H\Omega_T=iH\\
	&\implies \Omega_T^{^-1}(-i)\Omega_T\Omega_T^{^-1}H\Omega_T=iH\\
	&\implies i\Omega_T^{^-1}\Omega_T\Omega_T^{^-1}H\Omega_T=iH\\
	&\implies i\Omega_T^{^-1}H\Omega_T=iH\\
	&\implies  \Omega_T^{^-1}H\Omega_T=H
\end{align}
Which is consistent with the commutation with the Hamiltonian(\ref{comm}). Therefore the use of anti-unitary operator necessitates the use of complex numbers, where the property of anti-linearity $\Omega_T^{^-1}(-i)\Omega_T=i$ permits a consistent definition of the important discrete symmetry of time reversal. 
\section{Analytic Continuation and Wick's rotation}
A greater insight into the behavior of a system can be gained by the mathematical procedure of analytical continuation\cite{wick1}\cite{wick0}\cite{wick2}\cite{wick3}. Wick's rotation is a rotation in the complex time plane where Euclidean time is substituted to Minkowskian time through the transformation $t\longrightarrow-it$. There is deep relation between the time evolution operator in QM $U=e^{-iHt}$ and the partition function of equilibrium statistical mechanics $Z=Tr(e^{-\beta H})$. This connection is made clear through Feynman's path integral formulation of QM\cite{feyn}\cite{feyn2}. Analytic continuation permits a mathematically well defined treatment of the path integrals of a quantum system which are generally delicate to handle and proved useful in studying low temperature behavior and tunneling phenomena\cite{wick2}\cite{wick4}. It is clear that such a deep connection wouldn't be possible without complex formalism of QM.
\section{Heisenberg's Uncertainty Principal}
Let's for simplicity consider $\Psi(x,t)$ be the one dimensional wave function of a quantum system. The Fourier transform of $\Psi$ (at fixed time) decomposes it in terms of the momentum components of the wave function :

\begin{equation}\label{key}
	\Psi(x,t)=\int \frac{dp}{2\pi}\bar{\Psi}(p,t)e^{-ipx}
\end{equation}

It is well known that measures of the supports of $\Psi$\cite{supp}, say $\Delta_x$ and its Fourier transform $\hat{\Psi}$, say $\Delta_p$, are inversely proportional. The extreme and nonphysical but illustrative example is given by Dirac's delta function, $\hat{\Psi}(p,t)=f(t)\delta(p-p_0)$ which has a vanishing support $\Delta_p\longrightarrow0$, for a particle with an exact momentum $p_0$. The wave function takes then the form $\Psi(x,t)=\frac{f(t)e^{-ip_0x}}{2\pi}$ with an infinite support $\Delta_x\longrightarrow\infty$. Thus, in general:
\begin{equation}\label{key}
\Delta_x\Delta_p\sim1	
\end{equation}

Therefore, a perfect knowledge of the momentum of the particle precludes any knowledge about its position and no concept of particle's path can be exist. This is the physical content of Heisenberg's uncertainty principal\cite{lan}. Complex numbers are essential to this effect through the definition of the Fourier transform as $\Psi(x,t)=\frac{f(t)e^{-ip_0x}}{2\pi}$ is a spatial wave (for fixed time) but gives no information about the position at time $t$ because $||\Psi||^2=1$ For all spacetime points $(x,t)$ .
\section{An Apparatus out of Anti-matter}

In His Noble lecture, Schwinger\cite{sch2} made a remarkable observation about the implication of complex numbers in QM. He drew intention to the mathematical equivalence, inherent in QM, of the two square roots of $-1$, i.e. $\pm i$. He linked this arbitrariness to the nature of the measurement apparatus. He ask the following question "\textit{What general property of any measurement apparatus is subject to our control, in principle, but offers only the choice of two alternatives?}". A macroscopic measurement apparatus can be made out of matter but in principal it can be constructed from anti-matter too!. Given the validity of the TCP theorem\cite{tcp} which asserts the equivalence of matter and anti-matter, Complex numbers are necessary in QM to exhibit this equivalence. 
\section*{CONCLUSION}
As shown above, Quantum Mechanics incorporates Complex Numbers by necessity and not just by convenience, with deep insight still to be acquired by investigating further the complex formalism of QM.

\end{document}